\documentclass[reprint]{revtex4-2}

\usepackage{hyperref}
\usepackage{graphicx}
\usepackage{dcolumn}
\usepackage{bm}
\usepackage[utf8]{inputenc}
\usepackage[T1]{fontenc}
\usepackage{mathptmx}
\usepackage{etoolbox}
\usepackage{multirow}
\usepackage{color}
\usepackage{xcolor}
\usepackage{amsmath}
\usepackage{graphicx}
\usepackage{amssymb}
\usepackage{textcomp}
\usepackage{float}
\usepackage{natbib}
\usepackage{ragged2e}
\usepackage{color,soul}
\usepackage{array}
\usepackage{dcolumn}
\usepackage{url}
\usepackage{multirow}
\usepackage{appendix}
\usepackage[normalem]{ulem}
\usepackage{soul}
\usepackage{caption}
\usepackage{subcaption}
\usepackage{ulem}
\usepackage{braket}

\begin{document}

\preprint{AIP/123-QED}

\title[]{Optical formation of ultracold NaK$_2$ ground state molecules}

\author{Baraa Shammout}
\email{shammout@iqo.uni-hannover.de}
\affiliation{Institut für Quantenoptik, Leibniz Universität Hannover, 30167 Hannover, Germany}

\author{Leon Karpa}
\affiliation{Institut für Quantenoptik, Leibniz Universität Hannover, 30167 Hannover, Germany}

\author{Silke Ospelkaus}
\affiliation{Institut für Quantenoptik, Leibniz Universität Hannover, 30167 Hannover, Germany}

\author{Eberhard Tiemann}
\affiliation{Institut für Quantenoptik, Leibniz Universität Hannover, 30167 Hannover, Germany}

\author{Olivier Dulieu }
\email{olivier.dulieu@universite-paris-saclay.fr}
\affiliation{Universit$\acute{\text e}$ Paris-Saclay, CNRS, Laboratoire Aim$\acute{\text e}$ Cotton, Orsay, 91400, France}

\date{\today}

\begin{abstract}
We study the rovibronic transitions in NaK$_2$ between its electronic ground state $1^2A'$ and its second excited state $3^2A'$, to identify possible pathways for the creation of ultracold ground-state triatomic molecules. Our methodology relies on the computation of potential energy surfaces and transition dipole moment surfaces for the relevant electronic states using \textit{ab initio} methods. Rovibrational energy levels and wave functions are determined using the discrete variable representation approach. A double-well structure of the potential energy surface is identified for both states, and the related transition strengths between the rovibrational levels are derived. Our calculations show that the formation of ultracold ground-state NaK$_2$ molecules is expected when starting from an excited electronic state of NaK$_2$, which can be created by photoassociation of NaK and K observed by optical means by Cao \textit{et al} (Phys. Rev. Lett. 2024, \textbf{132}, 093403).
\end{abstract}
 \maketitle

\section{Introduction}
\label{sec:intro}

Ultracold chemistry, a vibrant research topic, has emerged due to remarkable experimental achievements in producing ultracold, dilute samples of diatomic bialkali molecules in their absolute ground state, both in homonuclear \cite{danzl2008} and heteronuclear (dipolar) \cite{ni2008} ensembles. It refers to the possibility of observing reactions between ultracold particles colliding at ultralow kinetic energies, $E \ll k_B\times 1$~mK (where $k_B$ is the Boltzmann constant), allowing to control these reactions with external electromagnetic fields. This long-sought goal is a major objective in the physical chemistry (or chemical physics) communities. Since the first instances of the term ''cold chemistry'' appearing in the titles of scientific literature,  \cite{weck2006,dulieu2006,krems2008}, numerous articles and books have been published to show the advancements made in this field \cite{balakrishnan2016,dulieu2017,bohn2017,heazlewood2019,perez-rios2021,softley2023,bohn2023}. An emblematic spontaneous chemical reaction in an ultracold gas is the one occurring between two KRb molecules prepared in their lowest ground state energy level, KRb+KRb $\rightarrow$ K$_2$+Rb$_2$, which is exothermic by about 10~cm$^{-1}$ \cite{ospelkaus2010a,ospelkaus2010b}. The formation of triatomic and tetra-atomic transient complexes induced by the so-called ''sticky collisions'' \cite{mayle2012,mayle2013} has been indirectly inferred later \cite{hu2019}. Due to the dipolar character of the reactants, this reaction can be controlled using an external electric field, revealing pronounced stereodynamics \cite{ni2010}. All bialkali dimers LiX (X = Na, K, Rb, Cs) prepared in the rovibronic ground state are expected to follow the same reaction, LiX+LiX $\rightarrow$ Li$_2$+X$_2$ \cite{zuchowski2010}. 

More recently, mixtures of ultracold atomic and molecular gases have been successfully created. These mixtures are likely to undergo chemical reactions of the type AB+A $\rightarrow$ A$_2$+B, even though the final products of these reactions have not yet been analyzed in experiments. As chemistry deals with the transformation of reactants and formation of new products, another opportunity is presented in the form of light-assisted ultracold collisions, such as AB+A $+ h\nu \rightarrow$ A$_2$B$^*$, where the colliding particles absorb a photon with suitable energy $h\nu$, to create a short-lived excited complex A$_2$B$^*$. This well-known process of photoassociation (PA) \cite{jones2006} was initially proposed to associate two cold atoms in a diatomic molecule (A+B+$h\nu \rightarrow$ AB$^*$). The excited complex AB$^*$ can decay into ultracold ground-state diatomic molecules, and this process has enabled the first observation of ultracold molecules ever \cite{fioretti1998}. Extensions of PA to atom-molecule systems have previously been discussed \cite{perez-rios2015,elkamshishy2022,shammout2023}, and recently realized in the mixture of K and NaK \cite{cao2024}.

Here, we investigate the possible decay of the excited complex A$_2$B$^*$ into ground-state A$_2$B molecules via spontaneous or stimulated emission. Following our previous work \cite{shammout2023}, we illustrate our study assuming the
successful PA of ultracold ground-state K atoms and NaK molecules. Our article is structured as follows. In Section \ref{sec:method}, we provide a description of optical transitions between the relevant molecular states. We present our methodology for electronic structure calculations yielding potential energy surfaces (PESs) and transition dipole moment (TDMs) surfaces, which we interpolate for the treatment of the nuclear motion to determine bound levels of the trimer. We present our results in Section \ref{sec:results} for the low-lying rovibrational levels of the electronic ground state $1A'$ and of the excited electronic state $3A'$, which are used to estimate the transition strengths between the ground state levels and several selected excited state levels. In Section \ref{sec:feasibility}, we discuss the experimental feasibility of the proposed scheme and find a promising perspective.

\begin{figure*}
    \centering
\includegraphics[scale=0.7]{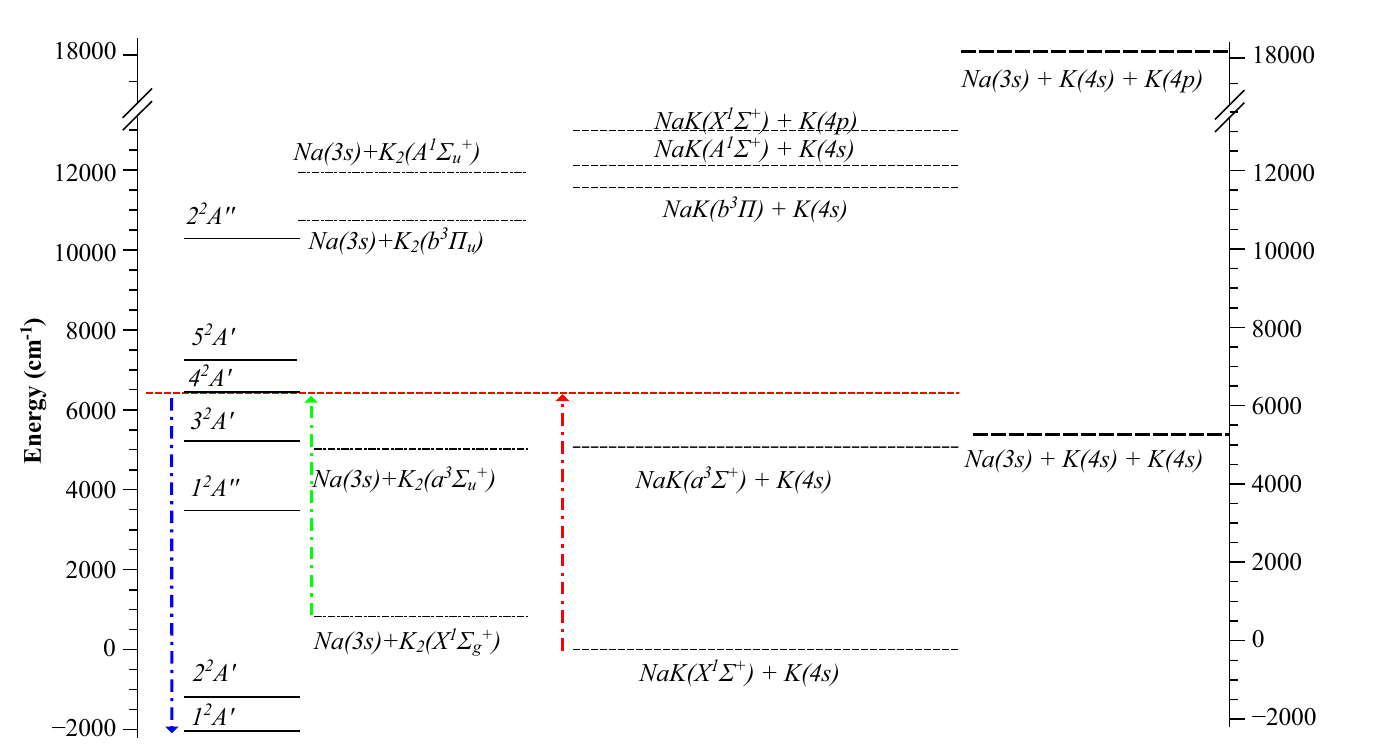}
    \caption{Energy diagram of the NaK$_2$ complex in various configurations: three separated atoms, atom + diatom arrangements, and triatom in the $C_s$ symmetry. The energy of the horizontal lines corresponds to the appropriate sums of atomic energies and of energies of the bottom of the relevant diatomic or triatomic PES. The reference of energies is taken at the initial state NaK($X^1\Sigma^+$) + K($4s$) of a PA experiment as in Ref. \onlinecite{cao2024} (red arrow). The same triatomic excited states could be reached by PA of Na($3s$) and K$_2$($X^1\Sigma_g^+$) (green arrow). The downward blue arrow illustrates the proposed transition for the formation of ultracold NaK$_2$ molecules in the absolute ground state.  }
    \label{fig:energy_diagram}
\end{figure*}

\section{Methodology}
\label{sec:method}

A simplified diagram of energy levels and relevant optical couplings is depicted in Fig. \ref{fig:energy_diagram}. A colliding pair of K and NaK absorbs a photon pictured by the upward red arrow referring to the photon energy used in Ref.\onlinecite{cao2024}. The reachable excited electronic state of NaK$_2$ corresponds to the transition from the $1^2A'$ electronic state to the $3^2A'$ electronic state in the notation of the $C_s$ point group. Thus, we will focus on the spontaneous or stimulated decay of the bound levels of the excited state $3^2A'$ down to the lowest bound levels of the $1^2A'$ ground state (downward blue arrow in Fig. \ref{fig:energy_diagram}).

\subsection{Electronic structure calculations}
\label{ssec:structure}

All \textit{ab initio} calculations were carried out with the software package MOLPRO~\cite{werner2012, werner2020} using the internally contracted multiconfiguration-reference configuration interaction (MRCI) method~\cite{werner1988, knowles1992} with Pople correction. We used effective core potentials of the Stuttgart–Cologne group ECP10SDF (Na) and ECP18SDF (K) together with the valence basis sets described in detail in our earlier work \cite{shammout2023} and the core polarization potentials (CPPs) of Ref.~\onlinecite{fuentealba1982}. Molecular orbitals were obtained from state-averaged MCSCF optimizations~\cite{werner1985} in $C_s$ symmetry with equal weights on the lowest three doublet $A'$ states and an active space of 7 orbitals (5 $A'$, 2 $A''$), which leads to a satisfactory convergence on the excited states of interest.

\subsubsection{Coordinate system}

\begin{figure}
\includegraphics[scale=0.68]{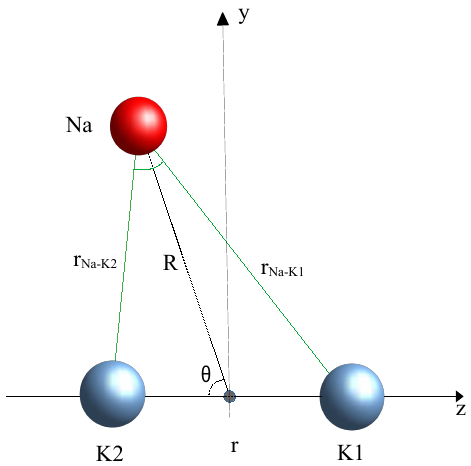}
    \caption{Geometry for the triatomic molecule NaK$_2$ with Jacobi coordinates ($R$, $r$, $\theta$) and two bond lengths and one bond angle coordinates ($r_{Na-K1}$, $r_{Na-K2}$ and $\angle$KNaK). The dipole moment $\Vec{d}$ can have a non-zero component in the direction of the $y$ and $z$ axes of the BF frame.} 
    \label{fig.coord}
\end{figure}

Based on the permutation symmetry of the identical atoms K in NaK$_2$, the system can be modeled within the $C_{2v}$ point group, or the $C_s$ point group in the general case. We calculate the PESs of the relevant electronic states in Jacobi coordinates ($R$, $r$, $\theta$) as depicted in Fig.\ref{fig.coord} with $r$ being the diatomic bond length of K$_2$ and $R$ the distance between the Na atom and the center of mass of K$_2$. The angle $\theta$ between the $\hat{R}$ and $\hat{r}$ axes only needs to be varied in the [$0^\circ-90^\circ$] range due to the permutation symmetry. For the triatomic molecule, placed in the $yz$ plane with the $x$-axis that points out of the plane (downward) to form a right-handed coordinate system, the $z$-axis aligns with the diatomic bond $r_{K-K}$, and the $y$-axis passes through the center of K$_2$ and is perpendicular to the $z$-axis. These axes define the body-fixed (BF) frame. The molecular dipole moment operator can have up to three non-zero components depending on geometry: two in-plane components along the $y$- and $z$-axes, and one out-of-plane component along the $x$-axis. The out-of-plane $x$-component transforms as $A''$ in $C_s$ symmetry and enables transitions between electronic states of opposite symmetry (i.e., $A' \leftrightarrow A''$), while the two in-plane components transform as $A'$ and enables transitions between electronic states of the same symmetry (i.e., $A' \leftrightarrow A'$). This coordinate framework allows us to use PESs of the relevant electronic states and the corresponding transition dipole moment surface in the calculations of nuclear motion and transition strengths, applying the DVR code mentioned later.

In the following, we study the PESs of the doublet electronic ground state and the two low-lying doublet excited electronic states. We calculate their equilibrium geometries and atomization energies and define accordingly the mesh grid appropriate for rovibrational-level calculations in the energy range of our interest. For visualization, we use bond lengths and bond angle coordinates ($r_{\textrm{Na-K1}}$, $r_{\textrm{Na-K2}}$ and $\angle$KNaK).

\subsubsection{Electronic states }

As part of a theoretical study on ultracold alkali-metal dimer–dimer reactions \cite{zuchowski2010}, Żuchowski and Hutson computed equilibrium geometries for the corresponding heteronuclear alkali-metal trimers. Their calculations showed that, depending on the constituent atoms, the absolute minimum of the ground-state PES corresponds either to a $C_{2v}$ symmetry (a $1^2A_1$ state, \textit{e.g.}, Cs$_2$Li, or a $1^2B_2$ state, e.g., K$_2$Li) or to a $C_s$ symmetry (a $1^2A'$ state, e.g., K$_2$Rb ). For readability, hereafter we will omit the''double'' label since we do not consider the quartet multiplicity of these species in this paper.

\begin{figure*}
\includegraphics[scale=0.7]{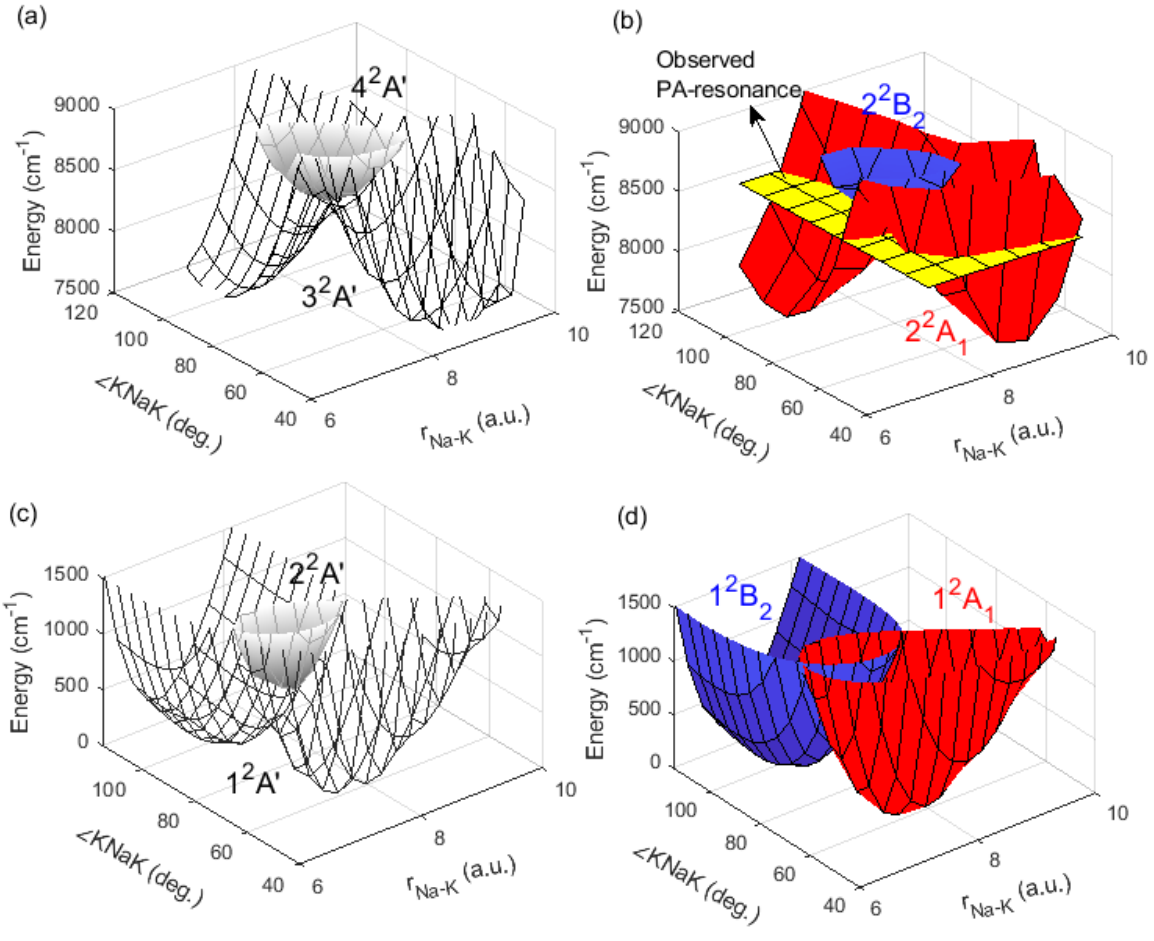}
    \caption{2-D cuts through the adiabatic PESs of NaK$_2$ doublet electronic states. (a) and (c): $1A'$, $2A'$, $3A'$ and $4A'$ PESs using a $C_s$ symmetry. (b) and (d): PESs of the corresponding states $1A_1$, $2A_1$, $1B_2$ and $2B_2$ using a $C_{2v}$ symmetry in bond length-bond angle coordinates ($r_{\textrm{Na-K}}$ and $\angle$KNaK). The region drawn in yellow corresponds to the energy range where PA resonances have been experimentally observed \cite{cao2024}. The origin of energies is taken at the minimum of the PES of the ground state $1A'$. }
 \label{Fig.minima_C2v}
\end{figure*}

Figure \ref{Fig.minima_C2v} shows 2D cuts of the PESs for the electronic states of interest, according to Fig. \ref{fig:energy_diagram}, plotted against the bond length $r_{Na-K}$ and the bond angle $\angle$KNaK coordinates. Panels (a) and (c) show calculations with $C_s$ symmetry constraint, resulting in states ($3A'$, $4A'$) and ($1A'$, $2A'$), respectively. Panels (b) and (d) display calculations performed in $C_{2v}$ symmetry, producing states labeled ($2A_1$, $2B_2$) and ($1A_1$, $1B_2$), respectively. 

In the $C_{2v}$ symmetry, the $1A_1$ and $1B_2$ PESs can cross due to the absence of symmetry-allowed coupling. However, when the symmetry constraint is relaxed to $C_s$, the PESs of states $1A'$ and $2A'$ exhibit a conical intersection seam \cite{liu2025, zuchowski2010, jasik2018}, which appears as an avoided crossing in this 2D slice. The minimum of $2A'$  PES lies in the energy region where $1A_1$ and $1B_2$ PESs intersect in $C_{2v}$ symmetry. Furthermore, we observe that the two minima depicted in panel (c) for the ground state PES are related to the minima of $1A_1$ and $1B_2$ PES in $C_{2v}$ and are well separated by a barrier of about 550 cm$^{-1}$. Consequently, they give rise to distinct vibrational structures, allowing us to neglect the influence of tunneling through the barrier on the corresponding vibrational levels at low energy. Similarly, the excited electronic states PESs $3A'$ and $4A'$ display an avoided crossing in $C_s$ symmetry, and the corresponding PESs $2A_1$ and $2B_2$ in $C_{2v}$ symmetry approach closely in energy at certain geometries but do not intersect. In particular, the PA resonances observed in the experiment by Cao \textit{et al.} \cite{cao2024} are located in the energy region of this $3A'/4A'$ avoided crossing. The electronic state assignment \cite{cao2024} for these resonances has been suggested to be $3A'(2A_1)$ and/or $1A'' (1B_1)$ \footnote{Ref.\protect\onlinecite{cao2024} assigned the observed resonances to the $1B_2$ and/or $3A'$ states in their $C_{2v}$ and $C_s$ symmetry labeling scheme. In contrast, using our coordinate conventions, the state they call $1B_2$ corresponds to the $1B_1$ state in our labeling, and the state they call $2B_1$ corresponds to the $2B_2$ state in our labeling. We believe that this difference may arise from the choice of symmetry axes and does not reflect a disagreement about the underlying physical state. }. In this work, we do not consider the $1A''$ state. Although the transitions $1A' \leftrightarrow 1A''$ are dipole allowed via the component perpendicular to the molecular plane, we have restricted our analysis to the $A'$ symmetry states for simplicity and clarity in this initial study. Including the $1A''$ state would require an additional set of potential energy and TDM surfaces along with expanded symmetry considerations, which is left for future work.

Table \ref{tab:eq_geom} summarizes the optimized stationary points of the PES of the ground state $1A'$ and the two excited low-lying  electronic states $2A'$, $3A'$  obtained from the MRCI calculations. For the ground state, we additionally performed spin‑unrestricted CCSD(T) \cite{knowles1993} calculations starting from restricted open‑shell Hartree‑Fock orbitals, denoted UCCSD(T)/ROHF with the same ECPs, CPPs and basis sets used for the MRCI calculations. The MRCI and UCCSD(T) calculations for the global minimum of ground-state PES are in good agreement with the results reported in Ref.~\onlinecite{zuchowski2010}. The global minimum has a distorted triangular geometry in the ground state PES $1A'$ as well as in the excited state PES $3A'$. The atomization energies are 7285~cm$^{-1}$ and 6433~cm$^{-1}$, relative to Na($3s$)+K($4s$)+K($4s$), for the states $1A'$ and $2A'$, respectively, while it is 13056~cm$^{-1}$, relative to Na($3s$)+K($4s$)+K($4p$), for the state $3A'$. The stationary points labeled ''saddle point'' and ''local minimum'' were derived by geometry optimizations in $C_{2v}$ symmetry under the constraint $r_{\mathrm{NaK}_1}=r_{\mathrm{NaK}_2}$. Subsequently, these geometries were confirmed in the PES $1A'$ and $3A'$ by single-point MRCI calculations in $C_s$ symmetry. Ref.~\onlinecite{zuchowski2010} did not explicitly report the locations of the local minimum or saddle point for $1A'$ PES, but provided the minima for the $1A_1$ and $1B_2$ PESs in symmetry $C_{2v}$ along with their atomization energies (see the supplementary material in Ref.\onlinecite{zuchowski2010}). When the symmetry restriction is relaxed to $C_s$ (allowing $r_{\mathrm{NaK}_1}\neq r_{\mathrm{NaK}_2}$), these two $C_{2v}$ structures develop into a first‑order saddle and a second minimum along the multidimensional $1A'$ PES that connects the distorted global $C_s$ minimum. By combining their reported atomization energies with ours, we estimate the relative energies of these structures and compare them in Table \ref{tab:eq_geom}. 

The saddle point of $1A'$ PES lies 38~cm$^{-1}$ above the global minimum (MRCI), which is consistent with the UCCSD(T) value 26~cm$^{-1}$ and the value 12~cm$^{-1}$ inferred from the data of Ref.~\onlinecite{zuchowski2010}. For the $3A'$ PES, the saddle point is 252~cm$^{-1}$ above its global minimum. The local minimum of $1A'$ PES is 140~cm$^{-1}$ above its global minimum in our MRCI results, also comparable to UCCSD(T) value (145~cm$^{-1}$) and close to the value inferred from Ref.~\onlinecite{zuchowski2010} (183~cm$^{-1}$). The local minimum of $3A'$ PES is 266~cm$^{-1}$ above its global minimum.

\begin{table*}[]
\caption{\label{tab:eq_geom} Optimized stationary‑point structures for NaK$_2$ in the electronic ground state $1A'$ and the two low-lying excited states $2A'$ and $3A'$ at the MRCI level of theory. For the ground state, a comparison with those calculated with the UCCSD(T) method as well as with the one of Ref.\onlinecite{zuchowski2010} is presented. T$_{e2}$ and T$_{e3}$ are the energies of the global minima of the two low-lying excited states relative to the energy of the global minimum of the ground state. Energies are referenced to the Na($3s$)+K($4s$)+K($4s$) asymptote for the $1A'$ and $2A'$ states, and to the Na($3s$)+K($4s$)+K($4p$) asymptote for the $3A'$ state. At the MRCI level of theory, the energy gap between the two asymptotes is 12965~cm$^{-1}$, and 5252~cm$^{-1}$ between the NaK(X$^1\Sigma^+$)+K($4s$) limit and the Na($3s$)+K($4s$)+K($4s$) limit (see Fig.\ref{fig:energy_diagram}).}
\begin{ruledtabular}
    
\begin{tabular}{cccccc}

Electronic state &
  Stationary point &
   Geometry $\&$ Energy &
  MRCI &
  UCCSD(T) &
   Ref.\onlinecite{zuchowski2010}\footnote[1]{The equilibrium geometry in Ref.\onlinecite{zuchowski2010} for the global minimum of the ground-state PES is given in terms of three bond lengths. Here we present it in terms of two bond lengths and bond angle. Data presented for the saddle point and local minimum correspond to the minimum of the $1A_1$ and $1B_2$ PESs, respectively in Ref.\onlinecite{zuchowski2010}.}   \\ \hline
\multirow{12}{*}{$1A'$} &
  \multirow{4}{*}{\begin{tabular}[c]{@{}c@{}}Global min. with\\ geometric symmetry $C_{s}$\end{tabular}} &
  $r_{\textrm{Na-K1}}$(a.u.) &
  7.06 &
  7.07 &
  6.99 \\
 &  & $r_\textrm{{Na-K2}}$(a.u.)     & 8.41   & 8.40  & 8.31  \\
 &  & $\angle$ KNaK ($^{\circ}$) & 58.79  & 58.52 & 58.42 \\
 &  & Energy (cm$^{-1}$)   & -7285   & -7178  & -7125  \\ \cline{2-6} 
 &
  \multirow{4}{*}{\begin{tabular}[c]{@{}c@{}} Saddle point with\\ geometric symmetry $C_{2v}$\end{tabular}} &
  $r_{\textrm{Na-K1}}$(a.u.) &
  7.67 &
  7.67 & 7.75
   \\
 &  & $r_{\textrm{Na-K2}}$(a.u.)     & 7.67   & 7.67  &  7.75     \\
 &  & $\angle$ KNaK ($^{\circ}$) & 58.32  & 58.20 &    58   \\
 &  & Energy (cm$^{-1}$)   & -7247   & -7152  &   -7113    \\ \cline{2-6} 
 &
  \multirow{4}{*}{\begin{tabular}[c]{@{}c@{}} Local min. with\\ geometric symmetry $C_{2v}$\end{tabular}} &
  $r_{\textrm{Na-K1}}$(a.u.) &
  7.00 &
  7.01 & 6.99
   \\
 &  & $r_{\textrm{Na-K2}}$(a.u.)     & 7.00   & 7.01  &   6.99    \\
 &  & $\angle$ KNaK ($^{\circ}$) & 94.98  & 93.77 &    98   \\
 &  & Energy (cm$^{-1}$)   &  -7145 \footnote[2]{At the MRCI level, the result with $C_{2v}$ basis is higher in energy by 58~cm$^{-1}$ than the one with $C_{s}$ basis, i.e. at -7087~cm$^{-1}$. The UCCSD(T) method shows a negligible difference (less than 1~cm$^{-1}$) in energy between the $C_{2v}$ and $C_{s}$ basis calculations.}   & -7033  &  -6942     \\ \hline
\multirow{5}{*}{$2A'$} &
  \multirow{5}{*}{\begin{tabular}[c]{@{}c@{}}Global min. with\\ geometric symmetry $C_{2v}$\end{tabular}} &
  $r_{\textrm{Na-K1}}$(a.u.)     & 7.26   &-&-  \\
 &  & $r_{\textrm{Na-K2}}$(a.u.) & 7.26   &-&-  \\
 &  & $\angle$ KNaK ($^{\circ}$) & 70.28  &-&-  \\
 &  & Energy (cm$^{-1}$)         & -6433   &-&-  \\
 &  & T$_{e2}$ (cm$^{-1}$)       & 852    &-&- \\ \hline
\multirow{13}{*}{$3A'$} &
  \multirow{5}{*}{\begin{tabular}[c]{@{}c@{}}Global min. with\\ geometric symmetry $C_{s}$\end{tabular}} &
  $r_{\textrm{Na-K1}}$(a.u.)     &  7.40  &-&- \\
 &  & $r_{\textrm{Na-K2}}$(a.u.) & 10.30  &-&-   \\
 &  & $\angle$ KNaK ($^{\circ}$) & 47.83  &-&- \\
 &  & Energy (cm$^{-1}$)         & -13056  &-&- \\
 &  & T$_{e3}$ (cm$^{-1}$)       & 7199   &-&- \\ 
 \cline{2-6} 
 &
  \multirow{4}{*}{\begin{tabular}[c]{@{}c@{}}Saddle point with\\ geometric symmetry $C_{2v}$\end{tabular}} &
  $r_{\textrm{Na-K1}}$(a.u.)     & 8.83  &-&-\\
 &  & $r_{\textrm{Na-K2}}$(a.u.) & 8.83  &-&-  \\
 &  & $\angle$ KNaK ($^{\circ}$) & 50.23 &-&- \\
 &  & Energy (cm$^{-1}$)         & -12804 &-&- \\ 
 \cline{2-6} 
 &
  \multirow{4}{*}{\begin{tabular}[c]{@{}c@{}}Local min. with\\ geometric symmetry $C_{2v}$\end{tabular}} &
  $r_{\textrm{Na-K1}}$(a.u.)     &  6.98 &-&- \\
 &  & $r_{\textrm{Na-K2}}$(a.u.) &  6.98 &-&- \\
 &  & $\angle$ KNaK ($^{\circ}$) & 103.96&-&- \\
 &  & Energy (cm$^{-1}$)         & -12790 &-&-  \\  
\end{tabular}
\end{ruledtabular}

\end{table*}

\begin{figure*}
\includegraphics[scale=0.7]{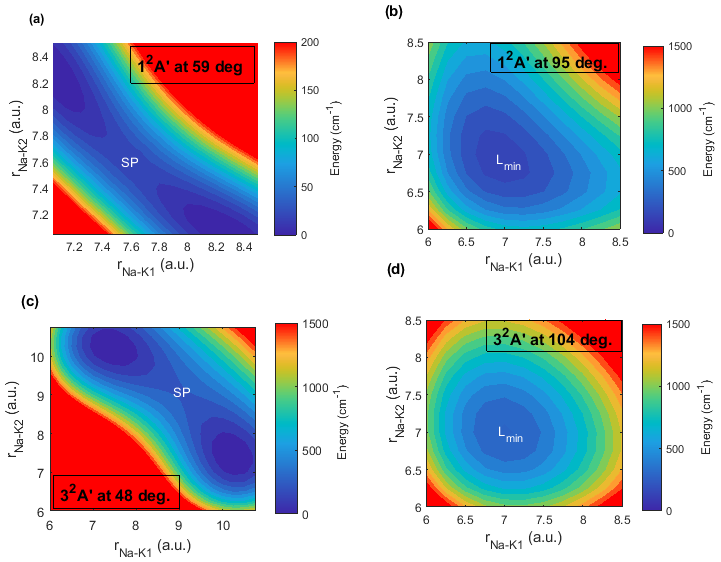}
    \caption{Adiabatic PESs of NaK$_2$ plotted in $(r_{\mathrm{NaK}_1},r_{\mathrm{NaK}_2})$ coordinates. Panels (a) and (b) show the ground $1A'$ state at fixed $\angle$KNaK bond angles of $59^\circ$ and $95^\circ$. Panels (c) and (d) show the excited $3A'$ state at $48^\circ$ and $104^\circ$. SP marks the $C_{2v}$-symmetric saddle-point structure; L$_{\min}$ marks the $C_{2v}$-symmetric local minimum. Energies are referenced to the global minimum of $1A'$ in (a,b) and to the global minimum of $3A'$ in (c,d). Panel (a) uses an expanded energy scale to make the saddle point visible.}
 \label{Fig.minima_contour}
\end{figure*}

Fig.\ref{Fig.minima_contour} shows a clear separation of the two minima in each of the $1A'$ [panels (a) and (b)] and $3A'$ [panels (c) and (d)] PESs. By distorting the isosceles acute triangle configuration (labeled ''SP'' in Fig.\ref{Fig.minima_contour}a,c) asymmetrically to a scalene triangle ($C_s$ symmetric structure) keeping the bond angle fixed, the electronic energies of the $1A'$ and $3A'$ states become lower than those of the corresponding saddle points. However, distorting the isosceles obtuse triangle configuration ($L_{min}$ in Fig.\ref{Fig.minima_contour} in (b) and (d)) to a scalene triangle ($C_s$ symmetric structure) while fixing $\angle$KNaK does not result in lowering energy. 

\subsubsection{Mesh grid for the PESs}
The geometries of the minima of the $1A'$ and $3A'$ PESs presented in Table \ref{tab:eq_geom} are expressed in terms of Jacobi coordinates ($R$, $r$, $\theta$) in Table \ref{tab:Jacobi_minima} to guide the choice of mesh design used in calculating the rovibrational eigenstates and the line strengths for the desired transitions. It should be noted that the equilibrium internuclear distance of the diatom K$_2$ in its ground state is $7.416$~a.u. \cite{pashov2008}. This value is close to the $r$ value found for the saddle-point structure of NaK$_2$ in the ground state, but differs significantly in the other geometries, which could indicate how much the adjacent Na atom perturbs the constituent diatom K$_2$.

\begin{table}[]
\caption{\label{tab:Jacobi_minima} Jacobi geometries for the stationary points of the PESs of the electronic ground state $1A'$ and the excited electronic state $3A'$ considered in our calculations in $C_s$ symmetry. }
\begin{ruledtabular}
\begin{tabular}{lllll}
\begin{tabular}[c]{@{}l@{}}Electronic \\ state\end{tabular} & Stationary point & $R$ (a.u.) & $r$ (a.u.) & $\theta$ ($^{\circ}$) \\ \cline{2-5} 
\multirow{3}{*}{$1A'$} & Global min.  & 6.75 & 7.69  & 78.4 and 101.6 \\
                     & Saddle point & 6.70  & 7.47  & 90             \\
                     & Local min.   & 4.73 & 10.32 & 90             \\ \cline{2-5} 
\multirow{3}{*}{$3A'$} & Global min.  & 8.11 & 7.65  & 65.6 and 114.4 \\
                     & Saddle point & 7.99 & 7.50   & 90             \\
                     & Local min.   & 4.30  & 11.00    & 90            
\end{tabular}
\end{ruledtabular}
\end{table}

The coordinates ($R$, $r$) are considered in the ranges $R\in$[2~a.u.-10~a.u.] and $r\in$[6~a.u.-14~a.u.] with a step size of 0.5~a.u. The Jacobi angle $\theta$ is taken between [40$^{\circ}$-90$^{\circ}$] with an increment of 5$^{\circ}$ so that the mesh grid consists of 3179 points. With this choice of grid points, the $3A'$ PES can be used to calculate rovibrational energies up to 1200~cm$^{-1}$ above its global minimum, covering the range where PA was observed, while the $1A'$ PES is used for rovibrational levels up to 350~cm$^{-1}$ above its minimum (see Section \ref{sec:results}.

\subsubsection{Transition dipole moment}

The electronic TDM between the $1A'$ and $3A'$ states as a function of the Jacobi coordinates is obtained from the matrix elements of the dipole operator on the electronic wave functions, calculated at the MRCI level. 

If we restrict the analysis to the $C_{2v}$ point group symmetry, two distinct transitions from the excited $2A_1$ state are allowed: one to the $1A_1$ state, which is driven solely by the $y$-component of the transition dipole moment (TDM), and another to the $1B_2$ state, which is governed by the $z$-component of the TDM. When the symmetry is reduced to $C_s$, both the $y$- and $z$-components of the TDM are active for the transition $3A' \rightarrow 1A'$ near the global and local minima.

Because we are interested in transitions towards the rovibrational levels localized around the global or local minima of the ground-state PES, we limit the TDM calculations to these two regions in the coordinate space. We choose the region around the global minimum  $R\in$[6.4~a.u.--8.75~a.u.],  $r\in$[7.7~a.u.--8.95~a.u.], $\theta\in$[67.5$^{\circ}$--89.2$^{\circ}$] with 13, 8 and 12 grid points, respectively (1248 grid points in total). The second region around the local minimum is taken as $R\in$[3.5~a.u.--5.75~a.u.], $r\in$[9.75~a.u.--11.5~a.u.], $\theta\in$[83.5~$^{\circ}$--89.5~$^{\circ}$] with 10, 8 and 4 grid points, respectively (320 grid points in total). These restrictions allow for reliable calculations of the TDM involving wave functions with a well-defined electronic character, far enough away from possible mixtures due to conical intersections.

The absolute values for the $y$- and $z$-components of the TDM $\vec{d}$ in the BF frame (defined in Fig.\ref{fig.coord}) for the $3^2A' \rightarrow 1^2A'$ transition at the geometry of the global minimum ($R=6.8$~a.u., $r=7.7$~a.u., $\theta=78^{\circ}$) of the $1A'$ PES are $|d_y|=1.19$~D and $|d_z|=0.89$~D, while $|d_z|=4.19$~D close to the geometry of the local minimum ($R=4.7$~a.u., $r=10.3$~a.u., $\theta=90^{\circ}$), which indicates an enhancement of the transition strength at the $C_{2v}$ geometry. A detailed description of this TDM surface is given in Appendix \ref{App:TDM}.

\subsection{Calculation of rovibrational levels and transition intensities}
\label{subsection: Calculation of rovibrational levels and transition intensities}

We selected the $^{23}$Na$^{39}$K isotopologue and the $^{39}$K isotope, which are the species used in the experiments in Hannover. We use the DVR3D program suite \cite{tennyson2004, tennyson2017}, which solves the Schrödinger equation for nuclear motion in a two-step variational procedure\cite{tennyson1986}, employing a discrete variable representation (DVR) for the wavefunctions in Jacobi coordinates ($R$, $r$, $\theta$) with Gauss-Laguerre quadrature for the integrals over $R$, $r$ and with Gauss-Legendre quadrature for the integrals over $\theta$. The angular basis functions are Legendre polynomials, while the radial basis functions are Morse oscillator-like functions with the variational parameters $r_e$, $R_e$ being effective equilibrium separations, $\omega_{R_e}$, $\omega_{r_e}$ effective frequencies related to the curvature of the potential, and $D_{R_e}$, $D_{r_e}$ related to dissociation energies. These parameters have been optimized to ensure convergence for the calculated rovibrational levels of the state $3A'$ and are reported in Table \ref{tab:morse} (more details on convergence can be found in Appendix~\ref{App:DVR_convergence}). The number of DVR points in $R$ and $r$ of the Gauss-Laguerre quadrature is set to 45, and the number of DVR points in $\theta$ of the Gauss-Legendre quadrature is set to 50 (see Table \ref{tab:DVR_grid} of Appendix \ref{App:DVR_convergence}). The same DVR grid and optimized Morse parameters are used for the state $1A'$, thus defining rovibrational wavefunctions on the same grid for both states. The \textit{ab initio} PESs and TDM components are evaluated on this grid by spline interpolation using the FORTRAN routine used in Ref.\onlinecite{klos2016}, adjusted to the desired geometry range, and the routines from Ref.~\onlinecite{prees1992}.

The full rovibrational basis in the DVR3D program (module ROTLEV3) is characterized by the quantum numbers ($J,M,n,p, q $), where $J$ corresponds to the total angular momentum and $M$ to its projection on the $Z$-axis of the space-fixed (SF) frame, $n$ is the index numbering rovibrational energies in increasing order, $p$ is associated with the rotational parity (as defined in DVR3D), with the total parity under inversion given by $(-1)^{J+p}$, where $p=0$ for $e$-parity states and $p=1$ for $f$-parity state, and $q$ is the parity associated with the permutation of identical nuclei in NaK$_2$ (even parity for $q=0$ and odd parity for $q=1$). 

\begin{table}[]
\caption{\label{tab:morse} Parameters for the Morse oscillator-like functions for the DVR calculations in $R$ and $r$. All values are in atomic units. }
\begin{ruledtabular}
\begin{tabular}{lll|lll}
$r_e$ & $\omega_{r_e}$           & $D_{r_e}$    & $R_e$ & $\omega_{R_e}$           & $D_{R_e}$  \\ \cline{1-6} 
10.08 & 1.27$\times 10^{-4}$ & 9$\times 10^{-3}$ & 6.098 & 1.38$\times 10^{-4}$ & 9$\times 10^{-3}$
\end{tabular}
\end{ruledtabular}
\end{table}

The strengths of the transitions between the rovibrational levels of the state $1A'$ (with double-primed labels) and of the state $3A'$ (with single-primed labels) are also produced by the DVR3D program (module DIPOLE3) with the TDM matrix elements in the SF frame being \cite{tennyson2004, zak2017}
 \begin{equation}
 T^{M'M''\tau}_{if}=\bra{ J'_{M'}, n', p', q'}\mu^s_\tau \ket{J''_{M''}, n'', p'', q'' },
\end{equation}
where $\mu^s_\tau$ is the $\tau$ component of the dipole moment operator $\Vec{\mu}$ in the SF frame resulting from the frame transformation (see Appendix A in Ref.\onlinecite{tennyson2004}) of the components $d_{z}(r, R, \theta)$ and $d_{y}(r, R, \theta)$ in the BF frame. 

The line strength is then given by
\begin{equation}
\label{Eq:line_strength}
    S(i-f)=\sum_{M'M''\tau}^{}(T^{M'M''\tau}_{if})^2.
\end{equation}
Note that the selection rules are: $\Delta J=0$ and $\Delta p=\pm 1$ ($e \leftrightarrow f$) or $\Delta J=\pm1$ and $\Delta p=0$ ($e\leftrightarrow e$ or $f\leftrightarrow f$). The permutation parity leads to the selection rule $\Delta q=\pm1$ (odd$\leftrightarrow$even) because the axis $z$ of the BF frame carries the coordinate $r$.

\section{Results and discussion}
\label{sec:results}

\subsection{Low-lying rovibrational levels of the electronic ground state $1A'$}
\label{subsec_vib_levels_ground}

\begin{figure*}
    \centering
\includegraphics[scale=0.65]{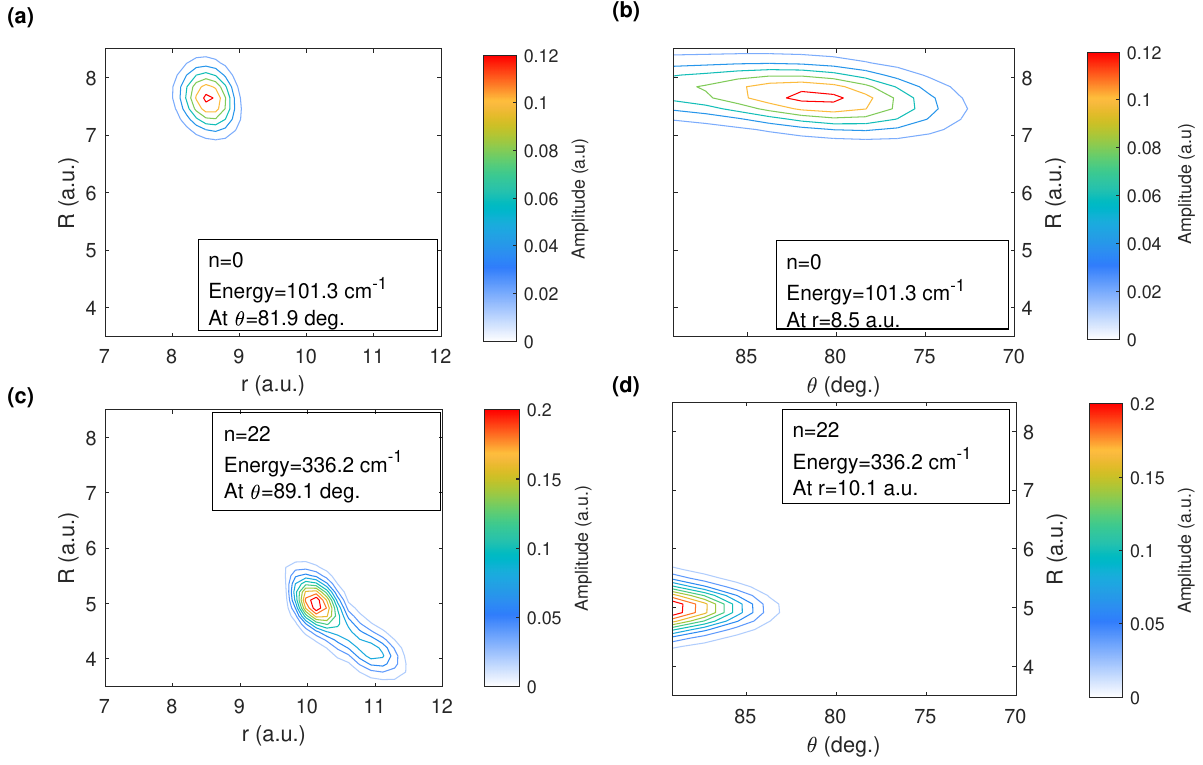}
    \caption{Contour plots in Jacobi coordinates ($R$, $r$, $\theta$) for wavefunctions of the lowest energy level $n=0$ (panels (a) and (b)) and of the level $n=22$ close to the local minimum (panels (c) and (d)) of the electronic ground state $1A'$. (a) at $\theta=81.9^{\circ}$; (c) at $\theta=89.1^{\circ}$; (b) at $r=8.5$~a.u.; (d) at $r=10.1$~a.u.. The colored lines reflect the amplitude of the wavefunction with the line spacing being 0.02~a.u.}
    \label{Fig.wf_ground}
\end{figure*}

\begin{figure*}
    \centering
\includegraphics[scale=0.60]{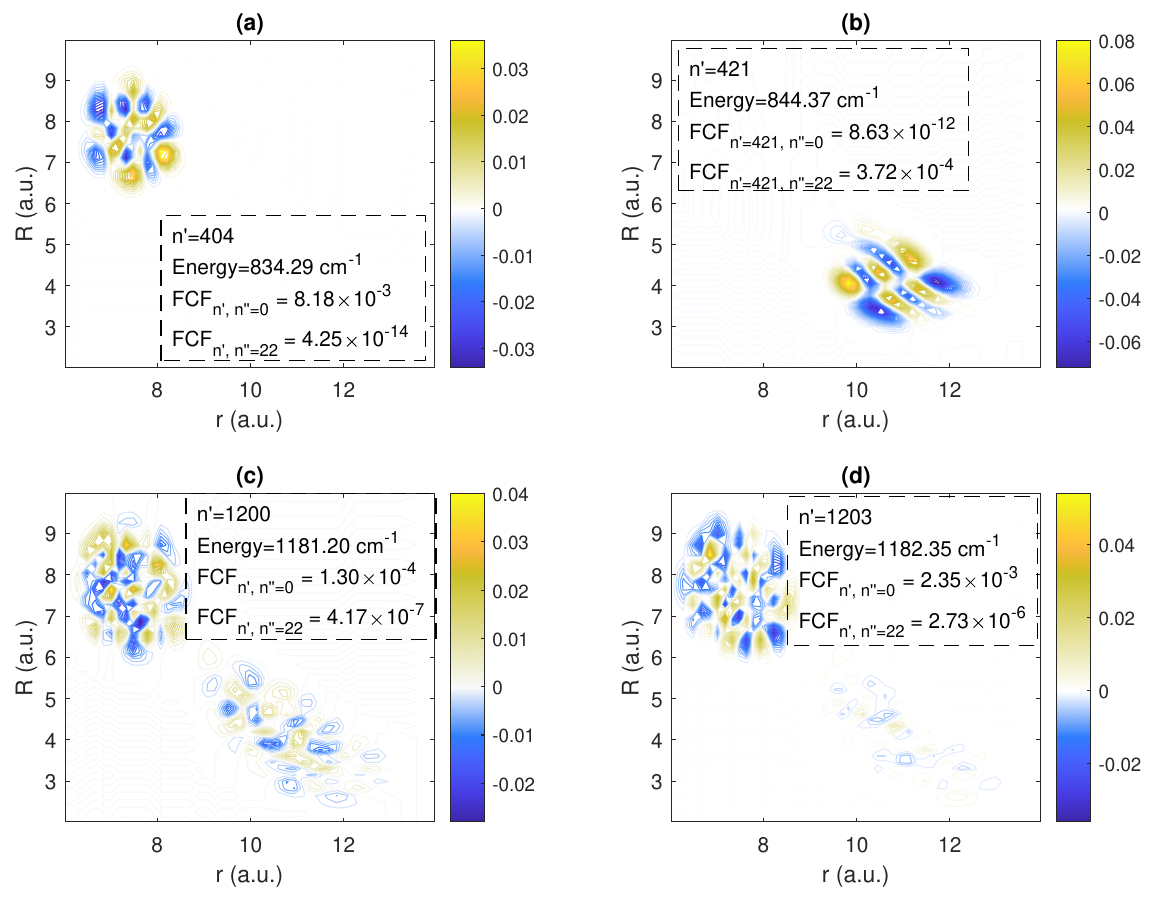}
    \caption{Contour plots  in Jacobi coordinates ($R$, $r$, $\theta=87.3^{\circ}$) for wavefunctions of selected energy levels $n'$ of the state $3A'$ (a) $n'=404$ at 834.29~cm$^{-1}$; (b) $n'=421$ at 844.37~cm$^{-1}$; (c) $n'=1200$ at 1181.2~cm$^{-1}$; (d)  $n'=1203$ at 1182.35~cm$^{-1}$. The origin of energy is at the global mininum of the PES. The values of the corresponding Franck-Condon factors FCF$_{n',n''}$ with the levels $n''=0$ and $n''=22$ of the state $1A'$ (see Fig. \ref{Fig.wf_ground}) are displayed. }
    \label{Fig.wf_selected_exci}
\end{figure*}

\begin{table}[]
\caption{ \label{tab:DVR-ground} The lowest 23 rovibrational levels for the state $1A'$ of $^{23}$Na$^{39}$K$_2$ calculated with the DVR3D method for $J=0$, $p=0$ and even permutation parity $q=0$. The zero of energy is taken at the global minimum of the $1A'$ PES.}
\begin{ruledtabular}
\begin{tabular}{lclc}

$n$ & Total energy (cm$^{-1}$) & $n$ & Total energy (cm$^{-1}$) \\ \cline{1-4}
0   & 101.33                   & 12  & 270.36                   \\
1   & 129.99                   & 13  & 275.17                   \\
2   & 163.54                   & 14  & 285.95                   \\
3   & 171.27                   & 15  & 296.45                   \\
4   & 195.27                   & 16  & 300.24                   \\
5   & 201.99                   & 17  & 307.31                   \\
6   & 214.82                   & 18  & 311.47                   \\
7   & 227.01                   & 19  & 312.60                   \\
8   & 231.44                   & 20  & 321.74                   \\
9   & 238.62                   & 21  & 332.89                   \\
10  & 261.32                   & 22  & 336.16                   \\
11  & 266.89                   &     &                                          
\end{tabular}
\end{ruledtabular}
\end{table}

In Tab.\ref{tab:DVR-ground} we present the low-lying rovibrational energy levels with $J=0$ for the state $1A'$ obtained from the DVR-calculation up to $n=22$. The zero point energy (ZPE) $n=0$ is found at 101.33~cm$^{-1}$ above the global minimum of the $1A'$ PES, thus also above the saddle point. The lowest rovibrational level localized around the local minimum is $n=22$ with a total energy of 336.16~cm$^{-1}$, that is 196~cm$^{-1}$ above the local minimum, which is the ZPE of the well-separated local minimum.   

The corresponding wavefunctions for the levels $n=0$ and $n=22$ are plotted in Fig.\ref{Fig.wf_ground} in Jacobi coordinates,  illustrating the significantly different vibrational structure around the local minimum compare to the one around the global minimum. As shown in the upper panel, the maximum amplitude of the ground vibrational wavefunction is at $\theta= 81.9~^{\circ}$. Although this might suggest a preference for a distorted geometry, considering the full symmetry by mirroring the right side reveals a maximum at $\theta= 98.1^{\circ}$ and thus the expected $C_{2v}$ symmetry. For the level $n=22$ localized around the local minimum (lower panel), the maximum amplitude of the wavefunction is indeed located at $\theta = 90^{\circ}$.

\subsection{Rovibrational levels for the excited electronic state $3A'$ }

The energy of the lowest level for the state $3A'$ is found at 7204.7~cm$^{-1}$ above the global minimum of the $1A'$ PES and 107~cm$^{-1}$ above the global minimum of the $3A'$ PES. We found 1253 rovibrational levels in the energy range up to 1200~cm$^{-1}$ with $J=0$, $p=0$ and $q=0$, while in Ref.\onlinecite{cao2024} the density of vibrational levels for the excited electronic state appears to be underestimated. There, around 600 vibrational levels were obtained within the same energy range according to a harmonic oscillator approximation. 

Figure~\ref{Fig.wf_selected_exci} presents contour plots for selected vibrational wavefunctions of the state $3A'$, corresponding to energy levels at 834.29~cm$^{-1}$, 844.37~cm$^{-1}$, 1181.20~cm$^{-1}$, and 1182.35~cm$^{-1}$ above the global minimum of the $3A'$ PES. The legend in each panel also includes the calculated Franck-Condon factors (FCFs) $\braket{ \psi(n') | \psi(n''=0)}^2$ and  $\braket{ \psi(n') | \psi(n''=22)}^2$ of these wavefunctions with the two vibrational wavefunctions of our special interest, namely $n''=0$ and $n''=22$ of the state $1A'$. Here, we can distinguish between three types of vibrational wavefunctions for the state $3A'$: 

\begin{itemize}
\item Type I, with wavefunctions localized around the global minimum of the $3A'$ PES (Fig.\ref{Fig.wf_selected_exci}a), for which the FCFs with the $n''=22$ wavefunction are very low, and the FCFs with the $n''=0$ wavefunction are relatively high.
\item Type II, with wavefunctions localized around the local minimum of the $3A'$ PES (Fig.\ref{Fig.wf_selected_exci}b). In contrast to type I, their FCFs with $n''=0$ are very low and relatively high with $n''=22$. States with wavefunctions of Type I and II appear at energies far below the barrier between the two wells of the $3A'$ PES. The clear separation of the two types is well illustrated by the two branches in the FCFs reported in Fig.\ref{Fig.FCF_Sif}(a). The data groups corresponding to the FCFs between the vibrational wavefunctions of the state $3A'$ and the vibrational
wavefunctions $n''= 0$ (black) and $n''= 22$ (red) of the $1A'$ both have one branch for high FCFs and one for low FCFs, respectively.
\item Type III, the wavefunctions can occupy (by tunneling through the barrier of the $3A'$ PES) both geometry regions around the global and local minima (Fig.\ref{Fig.wf_selected_exci}c,d). These vibrational wavefunctions appear at energies very close to the top of the barrier (1238~cm$^{-1}$ above the global minimum of the $3A'$ PES). In this case, both FCFs become relatively high and closer to each other (Fig.\ref{Fig.FCF_Sif}a) where the red and black data overlap.
\end{itemize}

\subsection{Line strengths }

\begin{figure}
    \centering
\includegraphics[scale=0.65]{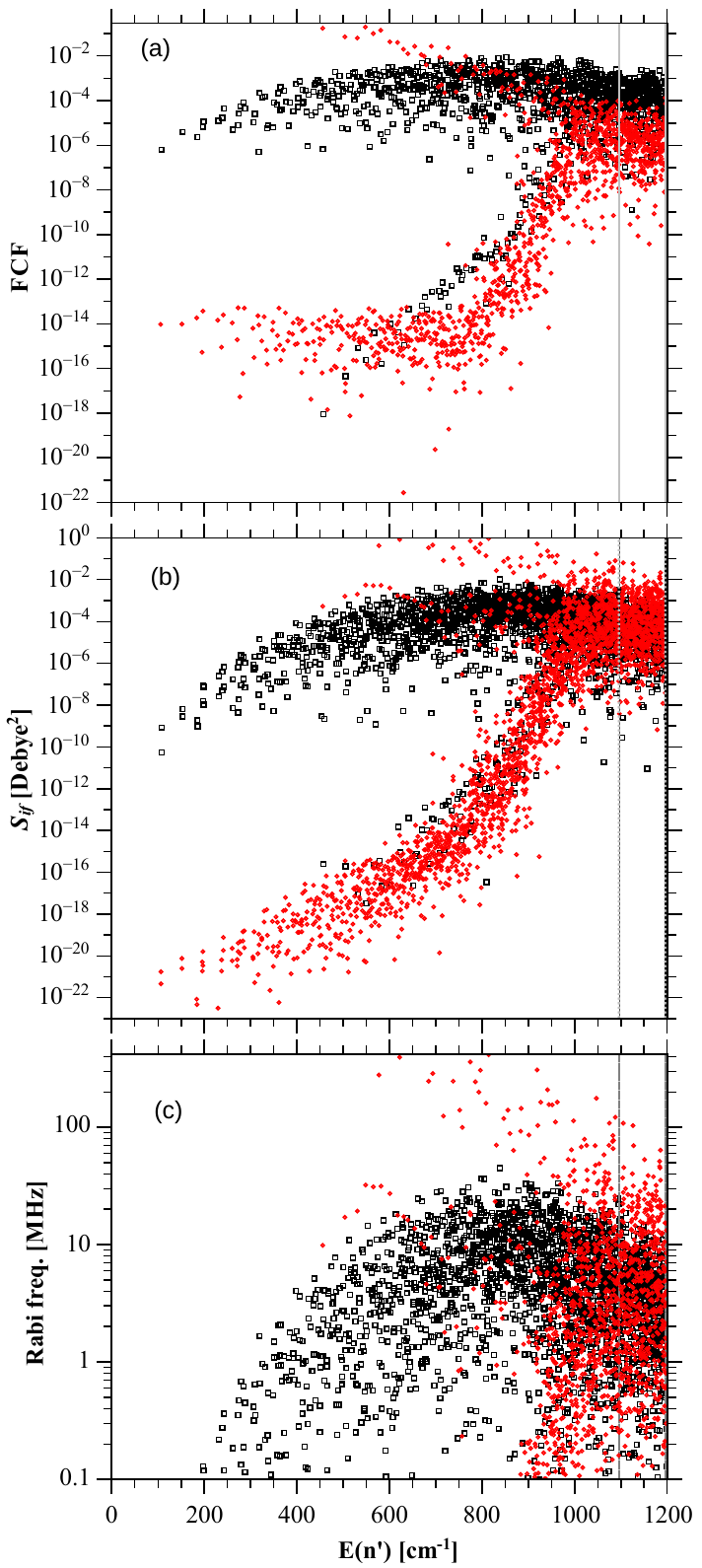}
    \caption{(a) Franck-Condon factors between vibrational wavefunctions of the excited electronic state $3A'$ (with energies $E(n')$) and vibrational wavefunctions $n'' = 0$ (black) and $n'' = 22$ (red) of the electronic ground state $1A'$. (b) Line strengths, expressed in [Debye]$^2$, corresponding to ro-vibronic transitions from the excited state $3A'$ ($J'=1$, $p'=0$, $q'=1$, $n'$) $\to$ $1A'$ ($J''=0$, $p''=0$, $q''=0$, $n''=0$ (black) and $n''=22$ (red)). (c) Rabi frequencies, in MHz, for the rovibronic transitions of (b), calculated at a laser intensity of 1~kW/cm$^2$. The transition energy is derived by E($n'$)-E($n''$)+T$_{e3}$ (see Table \ref{tab:eq_geom}). Vertical dashed lines indicate the experimentally observed positions of PA resonances  reported in Ref. \onlinecite{cao2024}.  }
    \label{Fig.FCF_Sif}
\end{figure}

Now, we will focus on transitions from the initial states $\ket{3A', J'=1, p'=0, q'=1, n'}$ to the desired final states $\ket{1A', J''=0, p''=0, q''=0, n''=0, 22}$, which could show the first estimate of the radiation decay distribution of an excited state from a PA process. The energy range for $n''$ from 0 to 22 is still small compared to the total transition energy. Thus, the power dependence on transition energy of the decay rate will have little influence on the distribution. Figure~\ref{Fig.FCF_Sif}b shows the calculated line strengths (in $D^2$) according to Eq.\ref{Eq:line_strength} (on a logarithmic scale) from the excited electronic state manifolds (ro-vibrational states with $J=1$, odd permutation parity and $e$-rotational parity) towards the ground vibrational level ($n''=0$, black squares) and towards the vibrational level localized around the local minimum ($n''=22$, red circles) of the electronic ground state up to an energy of 1200~cm$^{-1}$ relative to the global minimum of the excited electronic state. The vertical dashed lines indicate the two energy regions (1094.64-1096.57~cm$^{-1}$ and 1194.78-1196.63~cm$^{-1}$), where PA resonances were observed \cite{cao2024}. Due to an intrinsic uncertainty of about 100~cm$^{-1}$ in the absolute energy scale of our \textit{ab initio} calculations, a direct, one-to-one correspondence between individual calculated transitions and the ten resonances observed within the narrow experimental windows (one with a width of 1.93~cm$^{-1}$ and one with a width of 1.85~cm$^{-1}$) \cite{cao2024} is not feasible. Our calculations, for the same energy windows, yield approximately 12 vibrational energy levels for $J=0$ and 19 rovibrational levels for $J=1$; therefore, there is no conflict between observation and our evaluation. 

Our next step involves the study of line strengths within an extended energy window, specifically broadening the experimental window of 1094.64~cm$^{-1}$--1096.57~cm$^{-1}$ by 100~cm$^{-1}$, resulting in a new range of 994.64~cm$^{-1}$-–1196.57~cm$^{-1}$. We identified 20 distinct calculated transitions to the final state $n''=22$ that exhibit exceptionally high line strengths that are higher than the average by at least one order of magnitude. The average line strength for all transitions in that extended energy window is $1.4 \times 10^{-3} \ \text{D}^2$, while the average for these 20 dominant transitions is $4.1 \times 10^{-2} \ \text{D}^2$. Their energy spacings range from $0.5$ to $38 \ \text{cm}^{-1}$, which are easily resolved by laser spectroscopy. In contrast to the $n'' = 22$ case, transitions to the $n''= 0$ ground rovibronic level exhibit significantly lower line strengths. The average line strength in the extended energy window is $2.1 \times 10^{-4} \ \text{D}^2$, nearly an order of magnitude smaller than that of the $n'' = 22$ transitions. From this distribution, only 7 transitions stand out with line strengths exceeding the average by at least one order of magnitude. Their average line strength is $2.9 \times 10^{-3} \ \text{D}^2$, still notably weaker than the mean of strong transitions to $n'' = 22$. Moreover, the density of such strong transitions is substantially lower (7 versus 20). The energy separations among these 7 transitions range from $1$ to $44 \ \text{cm}^{-1}$.

\section{Experimental feasibility}
\label{sec:feasibility}

An efficient STIRAP (Stimulated Raman Adiabatic Passage) process to create ultracold NaK$_2$ molecules in the lowest vibrational level of the global and local minima ($n''$=0 or $n''$=22) relies on a sufficiently large two-photon Rabi frequency $\Omega$. This frequency is proportional to the Rabi frequencies of the two individual transitions ($\Omega_1$ for the Feshbach state to the excited state \cite{cao2024} and $\Omega_2$ for the excited state to the targeted ground state) and inversely proportional to the one-photon detuning $\Delta$. Maximizing $\Omega$ requires strong couplings ($\Omega_1$, $\Omega_2$) and manageable detuning. We now evaluate the Rabi frequencies ($\Omega_2$) of the transitions from levels $n'$ of the $3A'$ state to the targeted vibrational levels ($n''$=0 and $n''$=22) of the $1A'$ state. We assume a laser intensity of 1~kW/cm$^2$, which is comparable to the intensities used in \onlinecite{cao2024}. The calculations (Fig.\ref{Fig.FCF_Sif}c) reveal that transitions to $n''=22$ exhibit Rabi frequencies approaching 100 MHz. However, transitions to $n''=0$ are an order of magnitude weaker, indicating a potential difference in STIRAP efficiency in producing trimers in these two vibrational states. Further analysis should include calculating the values of $\Omega_1$  and optimizing the detuning $\Delta$ to determine the most effective STIRAP parameters.  

In the experiment of Ref.\onlinecite{cao2024}, two distinct pump lasers with wavelengths of 1558~nm and 1583~nm were used separately to excite weakly bound K$\cdots$NaK Feshbach molecules to intermediate electronic states located in two different spectral regions. Based on our \textit{ab initio} and DVR calculations, the energy separation between these intermediate states and the targeted rovibrational levels ($n'' = 0$ and $n'' = 22$) is estimated with an uncertainty of approximately 100~cm$^{-1}$. This corresponds to a wavelength uncertainty of roughly 15~nm for transition energies near 8350~cm$^{-1}$ (resulting in a 1198~nm Stokes wavelength when using the 1558~nm pump) and 8249~cm$^{-1}$ (yielding a 1212~nm Stokes wavelength with the 1583~nm pump), both targeting the $n'' = 0$ state. 

Since transitions to the local minimum configuration ($n'' = 22$ in our labeling) have much higher line strengths and Rabi frequencies than to the absolute ground vibrational state ($n'' = 0$), initial experiments might succeed faster if they first target the vibrational level in the local minimum configuration, then later try to address the absolute ground state in the global minimum configuration.

\section{Conclusion}

In conclusion, this study has explored the feasibility of creating ultracold NaK$_2$ triatomic molecules in their ground state. Through \textit{ab initio} calculations, we have mapped the relevant PESs and transition dipole moments, identifying a double-well structure in both the ground state $1A'$ and the excited state $3A'$. Our analysis of vibrational energy levels and transition strengths suggests that PA from an initial (NaK--K) complex, and also probably from a (K$_2$--Na) complex, to a bound level of the excited state $3A'$, followed by radiative decay to the ground state, offers a promising pathway for trimer formation. We have shown that transitions to both the global minimum and the local minimum configurations of the ground state are possible, opening the door to the creation of ground-state NaK$_2$ trimers with vibrational states associated with distinct structural configurations. We identified suitable Stokes laser wavelengths in the infrared region to drive these transitions in a STIRAP process and estimated the associated Rabi frequencies, based on the experimental work of Ref. \onlinecite{cao2024}. Specifically, to reach the vibronic ground state, a Stokes laser with wavelength 1198(15)~nm is required for a pump laser with wavelength 1558~nm, while transitions to the vibrational level near the local minimum of the $1A'$ PES promise a higher transfer efficiency. Our calculations suggest that these transitions will occur at wavelengths detuned by about $\sim$34~nm to the red from the transitions toward $n''=0$. 

These findings pave the way for future experimental investigations aimed at the creation of ultracold ground-state NaK$_2$ molecules and the exploration of their unique properties in the ultracold regime. On the theory side, the influence of the conical intersection between the $1A'$ and $2A'$ states on the PA dynamics, as well as the contribution of the $A''$ states to the PA transitions, will have to be modeled.

\section{Acknowledgments}
B.S., L.K. and S.O. gratefully acknowledge financial support from the Deutsche Forschungsgemeinschaft (DFG, German Research Foundation) through CRC 1227 (DQ-mat), Project No. A03, and under Germany’s Excellence Strategy - EXC-2123 QuantumFrontiers - 390837967, and the European Research Council through ERC Consolidator Grant No. 101045075 - TRITRAMO.
\newpage
 \clearpage
 \setcounter{figure}{0}
 \renewcommand{\thefigure}{A\arabic{figure}}

\section{Appendix}

\subsection{Convergence of the DVR calculations}
\label{App:DVR_convergence}
Reaching convergence in DVR calculations requires careful tuning of several control parameters. This includes optimization of the Morse oscillator-like basis functions and the size of the basis set, specifically the number of DVR points along each coordinate $R$, $r$, and $\theta$. For each radial coordinate $R$ and $r$, a set of Morse parameters must be defined: $R_e$, $\omega_{R_e}$, $D_{R_e}$ for $R$, and $r_e$, $\omega_{r_e}$, $D_{r_e}$ for $r$. The main goal is to determine parameter values that minimize errors in the computed vibrational energy levels, with special attention to the upper energy region. The values of $D_{R_e}$ and $D_{r_e}$ are found \cite{tennyson2004} to have a minor influence, allowing for some flexibility in their selection. Initial estimates for $R_e$ and $r_e$ are typically obtained from equilibrium bond distances of the PES, although the optimal values often deviate from these reference points. The initial guesses for $\omega_{R_e}$ and $\omega_{r_e}$ are usually based on the curvature of the PES. Importantly, $R_e$ is strongly coupled with $\omega_{R_e}$, just as $r_e$ is with $\omega_{r_e}$, and effective optimization requires adjusting these pairs simultaneously. To broaden the basis set and better accommodate highly excited vibrational states, it is common to increase $R_e$ and $r_e$ while reducing $\omega_{R_e}$ and $\omega_{r_e}$. Our PES for the electronic ground state $1A'$ and for the excited state $3A'$ present a particular challenge due to the presence of two distinct minima: one at $R = 8.11$ a.u., $r = 7.65$ a.u., and another at $R = 4.30$ a.u., $r = 11.0$ a.u. for the excited state $3A'$. This implies that the system can explore a wide range of $R$ and $r$ values, necessitating a flexible basis set and an extended DVR grid to accurately represent the wavefunction across both regions. To address this, we initially selected intermediate values for $R_e$ and $r_e$ that lie between the two minima. Specifically, we chose:
$R_e = (8.11 + 4.30)/2 = 6.21$ a.u. and 
$r_e = (7.65 + 11.0)/2 = 9.33$ a.u.. This provides a balanced starting point for capturing the essential physics of both potential wells. The optimization process was carried out iteratively. We began with a relatively small DVR grid and basis set to limit computational cost in the early stages. A grid of parameter combinations was then systematically generated, and DVR3D calculations were performed for each set. A key aspect of this process is ensuring convergence not merely producing a large number of eigenvalues, but ensuring that these eigenvalues remain stable with respect to changes in the DVR grid and basis size. When a specific set of Morse parameters results in little scatter of eigenvalues within the target energy range, it indicates that the basis set is effectively spanning the relevant regions of the PES and accurately representing the vibrational states.

\begin{figure}[]
    \centering
\includegraphics[scale=0.65]{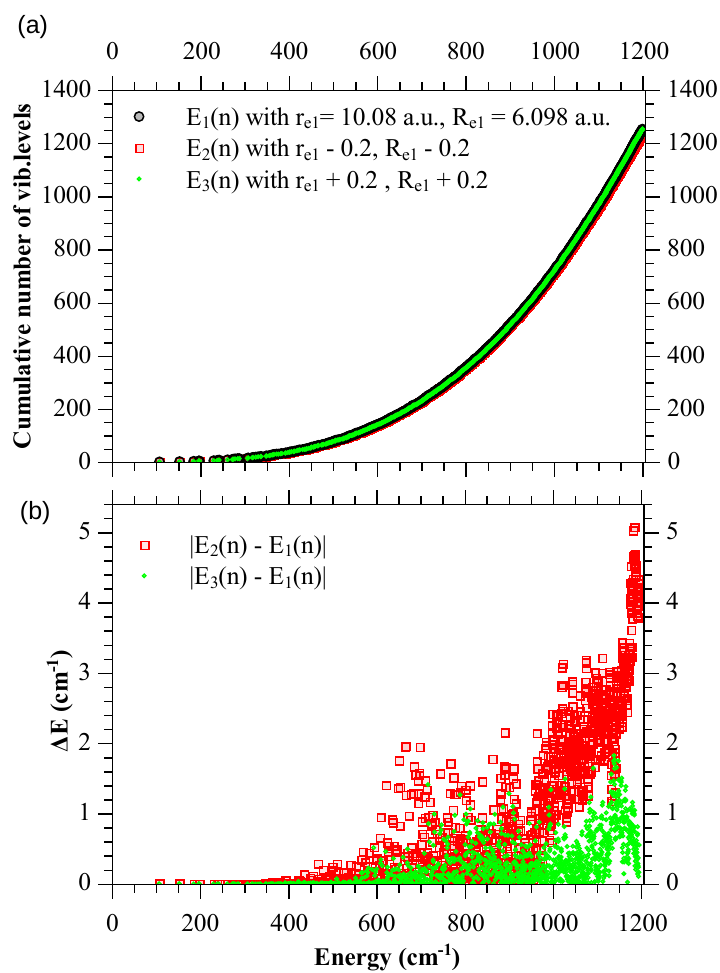}
    \caption{Stability of the vibrational levels for the excited electronic state $3A'$ calculated using the DVR method for $J=0$. (a) Cumulative number of vibrational levels plotted as a function of energy from the minimum of the PES, up to 1200~cm$^{-1}$ above it, showing the effect of tuning the optimized Morse parameters $R_{e1}=6.098$~a.u. and $r_{e1}=10.08$~a.u. (energies $E_1(n)$, black circles) by -0.2~a.u. (energies $E_2(n)$, red squares) and +0.2~a.u. (energies $E_3(n)$, green circles). (b) Absolute energy deviations $\Delta E(n)$ in cm$^{-1}$ from the energy levels $E_1(n)$ with optimized Morse parameters. Red squares: $\Delta E(n) = |E_2(n) - E_1(n)|$. Green circles: $\Delta E(n) = |E_3(n) - E_1(n)|$. }
    \label{Fig.convergence_morse}
\end{figure}

\begin{figure}[]
    \centering
\includegraphics[scale=0.65]{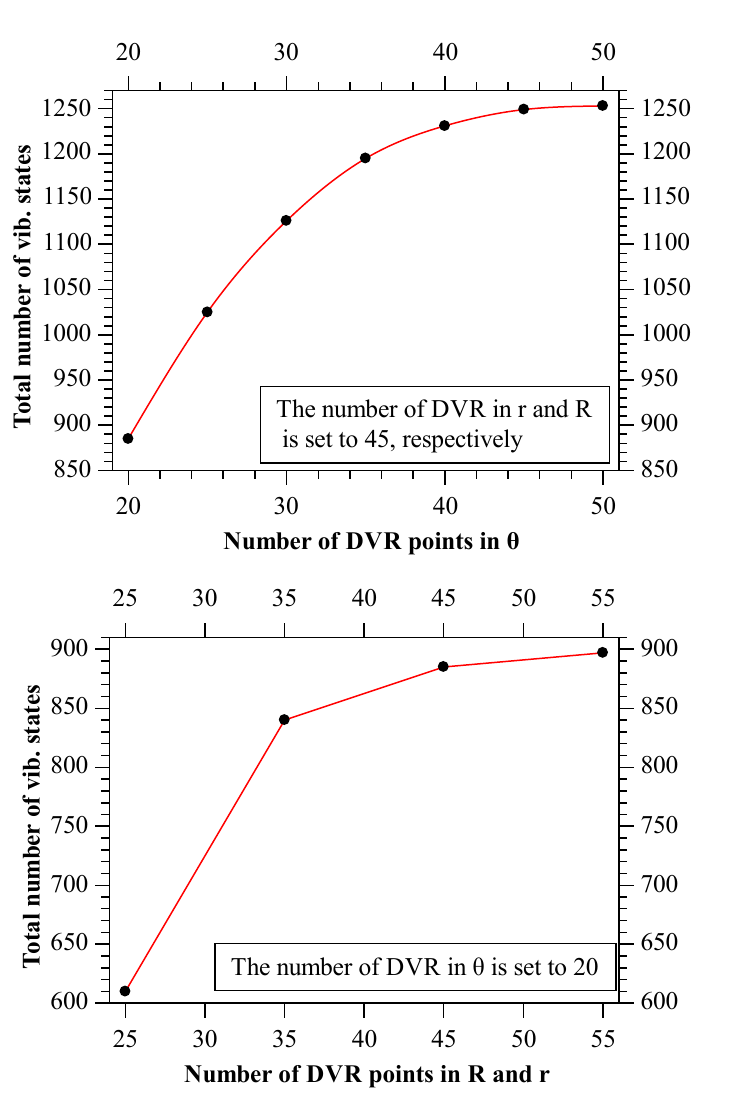}
    \caption{ Convergence of the total number of vibrational levels of the $J=0$ excited state $3A'$ up to 1200~cm$^{-1}$ as function of the number of the DVR points in $\theta$ (upper panel) and in $R$ and $r$ (lower panel). }
    \label{Fig.convergence_DVR_number}
\end{figure}

The final optimized Morse parameters are provided in Table \ref{tab:morse}. As shown in Figure \ref{Fig.convergence_morse}, the computed vibrational energy levels up to 1200 cm$^{-1}$ above the global minimum of the excited electronic state exhibit excellent stability even when $R_e$ and $r_e$ are varied by $\pm 0.2$~a.u. This robustness is demonstrated at the final DVR grid size, which uses 45 DVR points in both $R$ and $r$, and 50 points in $\theta$ (see Table \ref{tab:DVR_grid}). Fig.\ref{Fig.convergence_DVR_number} shows the number of DVR points required to achieve a stable eigenstate spectrum.

 \begin{table}[]
\caption{\label{tab:DVR_grid} Gauss-Laguerre quadrature points ($r_{\textrm{DVR}}$, $R_{\textrm{DVR}}$) for the Jacobi coordinates $r$ and $R$,  and Gauss-Legendre quadrature points $\theta_{\textrm{DVR}}$ for the angular coordinate $\theta$ used in the DVR calculations.}
\begin{ruledtabular}
\begin{tabular}{lll}
$r_{\textrm{DVR}}$(a.u.) & $R_{\textrm{DVR}}$(a.u.) & $\theta_{\textrm{DVR}}$ ($^{\circ}$) \\
\hline
13.95242     & 9.96065       & 1.37100           \\
13.56603     & 9.57017       & 3.14702           \\
13.25114     & 9.25222       & 4.93353           \\
12.97379     & 8.97237       & 6.72241           \\
12.72091     & 8.71737       & 8.51219           \\
12.48576     & 8.48038       & 10.30240          \\
12.26424     & 8.25725       & 12.09286          \\
12.05365     & 8.04524       & 13.88346          \\
11.85207     & 7.84241       & 15.67417          \\
11.65810     & 7.64731       & 17.46494          \\
11.47064     & 7.45885       & 19.25576          \\
11.28882     & 7.27615       & 21.04662          \\
11.11195     & 7.09849       & 22.83750          \\
10.93943     & 6.92528       & 24.62841          \\
10.77079     & 6.75603       & 26.41934          \\
10.60560     & 6.59030       & 28.21028          \\
10.44349     & 6.42774       & 30.00123          \\
10.28415     & 6.26800       & 31.79219          \\
10.12727     & 6.11080       & 33.58316          \\
9.97261      & 5.95588       & 35.37413          \\
9.81992      & 5.80298       & 37.16511          \\
9.66897      & 5.65189       & 38.95610          \\
9.51956      & 5.50239       & 40.74709          \\
9.37150      & 5.35428       & 42.53809          \\
9.22457      & 5.20737       & 44.32908          \\
9.07860      & 5.06146       & 46.12008          \\
8.93340      & 4.91637       & 47.91109          \\
8.78878      & 4.77191       & 49.70209          \\
8.64454      & 4.62787       & 51.49310          \\
8.50047      & 4.48407       & 53.28411          \\
8.35637      & 4.34027       & 55.07512          \\
8.21198      & 4.19624       & 56.86613          \\
8.06706      & 4.05172       & 58.65715          \\
7.92131      & 3.90642       & 60.44816          \\
7.77438      & 3.76000       & 62.23918          \\
7.62588      & 3.61206       & 64.03020          \\
7.47531      & 3.46212       & 65.82121          \\
7.32208      & 3.30958       & 67.61223          \\
7.16541      & 3.15366       & 69.40325          \\
7.00423      & 2.99332       & 71.19427          \\
6.83706      & 2.82709       & 72.98529          \\
6.66166      & 2.65273       & 74.77631          \\
6.47427      & 2.46653       & 76.56733          \\
6.26758      & 2.26124       & 78.35836          \\
6.02235      & 2.01779       & 80.14938          \\
             &               & 81.94040          \\
             &               & 83.73142          \\
             &               & 85.52244          \\
             &               & 87.31347          \\
             &               & 89.10449         
\end{tabular}
\end{ruledtabular}
\end{table}

\subsection{TDM surfaces}
\label{App:TDM}
\begin{figure*}
    \centering
\includegraphics[scale=0.60]{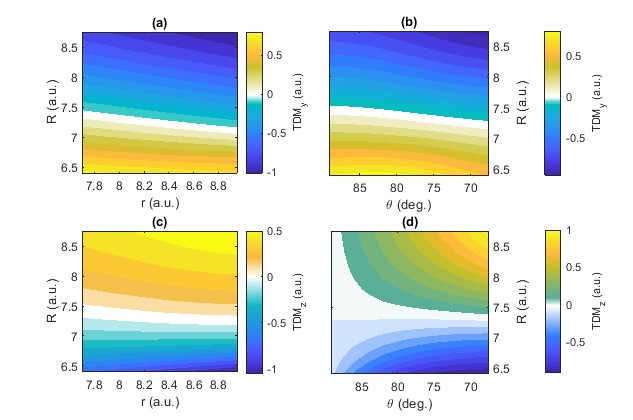}
   \caption{Contour plots in the range of the global minimum of $1A'$ for the transition dipole moment components $d_y$ (upper panel) and $d_z$ (lower panel) between the ground state $1A'$ and the excited state $3A'$ in Jacobi coordinates ($R$,$r$,$\theta$) at the fixed angle $\theta=80^\circ$ in (a) and (c) and at $r=7.7$~a.u. in (b) and (d). Line spacing is 0.1~a.u.}
    \label{Fig.TDM_global}
\end{figure*}

Figure \ref{Fig.TDM_global} illustrates the variation of the transition dipole moment (TDM) components $d_y$ and $d_z$ for the $3^2A' \rightarrow 1^2A'$ transition in the vicinity of the global minimum of the electronic ground state, expressed in Jacobi coordinates. Panels (a) and (c) display the TDM components as functions of $R$ and $r$ at a fixed Jacobi angle of $\theta = 80^\circ$, while panels (b) and (d) show the variations as functions of $R$ and $\theta$ at a fixed diatomic bond length of $r_{\mathrm{K-K}} = 7.7$~a.u.. At the $C_{2v}$ geometry ($\theta = 90^\circ$), the $z$-component of the TDM vanishes, consistent with the dipole selection rule that only the y-component can induce the $2^2A_1 \rightarrow 1^2A_1$ transition.

\begin{figure*}
    \centering
\includegraphics[scale=0.80]{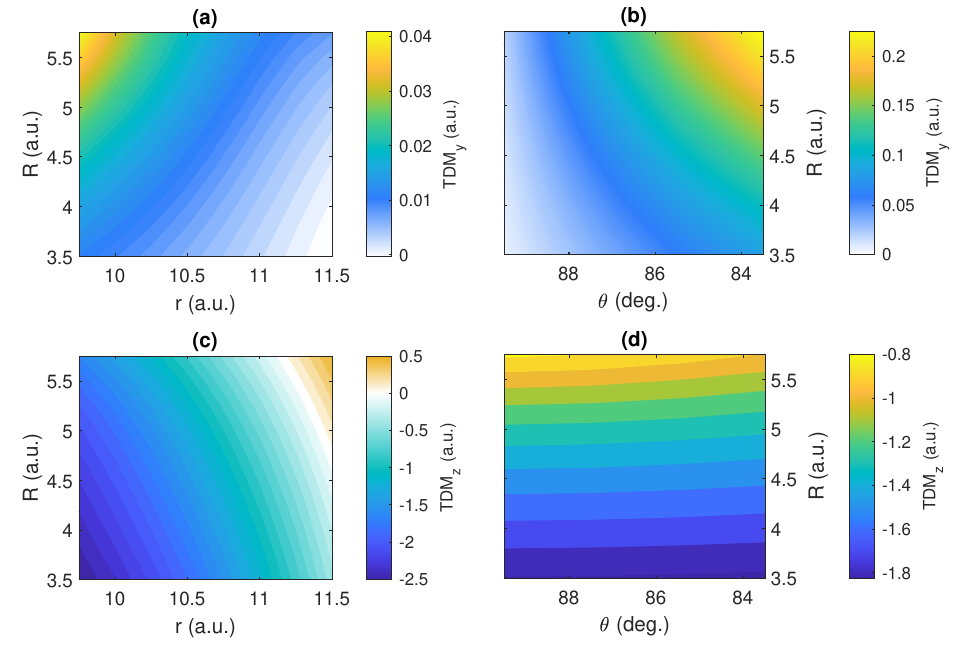}
   \caption{Contour plots in the range of the local minimum of $1A'$ for the transition dipole moment components $d_y$ (upper panel) and $d_z$ (lower panel) between the ground state $1A'$ and the excited state $3A'$ in Jacobi coordinates ($R$,$r$,$\theta$) at the fixed angle $\theta=89.5^\circ$in (a) and (c) and at $r=10.5$~a.u. in (b) and (d). Line spacing is 0.1~a.u. for $d_z$ and $10^{-3}$~a.u. for $d_y$. }
    \label{Fig.TDM_local}
\end{figure*}

In Figure \ref{Fig.TDM_local}, we present the TDM components in the region of the local minimum of the ground state PES. The left panels (a) and (c) correspond to a fixed Jacobi angle of $\theta = 89.5^\circ$, and the right panels (b) and (d) show the behavior at a fixed bond length of $r_{\mathrm{K-K}} = 10.5$ a.u.. In this region, the $y$-component of the TDM approaches zero as $\theta$ approaches $90^\circ$, according to the selection rule in $C_{2v}$ symmetry, where the transition $2^2B_2 \rightarrow 1^2A_1$ is allowed only via the $z$-component.

\newpage
 \clearpage

	\bibliographystyle{achemso}
	\bibliography{bibliocold,bibnote}

\end{document}